\documentclass[10pt,twocolumn,letterpaper]{article}

\usepackage{iccv}
\usepackage{times}
\usepackage{epsfig}
\usepackage{graphicx}
\usepackage[algo2e]{algorithm2e}
\usepackage{algorithm}
\usepackage{booktabs}
\usepackage{multirow}
\usepackage{color}
\usepackage{setspace}
\usepackage{amsmath}
\usepackage{amssymb}
\usepackage{microtype}
\usepackage[normalem]{ulem}
\usepackage{tikz}
\usepackage[accsupp]{axessibility}
\usepackage{comment}
\usepackage{colortbl}
\usepackage[skip=3pt]{caption}

\definecolor{gtgray}{gray}{0.97}
\definecolor{mygray}{gray}{.88}

\definecolor{gray1}{gray}{.90}
\definecolor{gray2}{gray}{.92}
\definecolor{gray3}{gray}{.94}
\usepackage[breaklinks=true,bookmarks=false]{hyperref}
\makeatletter
\def\hlinew#1{%
  \noalign{\ifnum0=`}\fi\hrule \@height #1 \futurelet
   \reserved@a\@xhline}
\makeatother

\usepackage[capitalize]{cleveref}
\crefname{section}{Sec.}{Secs.}
\Crefname{section}{Section}{Sections}
\Crefname{table}{Table}{Tables}
\crefname{table}{Tab.}{Tabs.}

\iccvfinalcopy % *** Uncomment this line for the final submission

\def\iccvPaperID{3218} % *** Enter the ICCV Paper ID here

\ificcvfinal\fi

\begin{document}
\twocolumn[{%
\renewcommand\twocolumn[1][]{#1}%
\title{Emotional Listener Portrait: Realistic Listener Motion Simulation in Conversation}
\author{Luchuan Song$^{1}$ $\quad$ Guojun Yin$^{2}$ $\quad$ Zhenchao Jin$^{3}$ $\quad$ Xiaoyi Dong$^{4}$ $\quad$ Chenliang Xu$^{1}$\\
$^{1}$University of Rochester $\quad$ $^{2}$University of Science and Technology of China\\
$^{3}$University of Hong Kong $\quad$ $^{4}$Shanghai AI Laboratory\\
{\tt\small \{lsong11@ur., chenliang.xu@\}rochester.edu, gjyin@mail.ustc.edu.cn,}\\
{\tt\small blwx96@connect.hku.hk, dongxiaoyi@pjlab.org.cn}
}
\maketitle
\thispagestyle{empty}
\begin{center}
    \centering
    \captionsetup{type=figure}
    \vspace{-0.7cm}
    \includegraphics[width=0.98\textwidth]{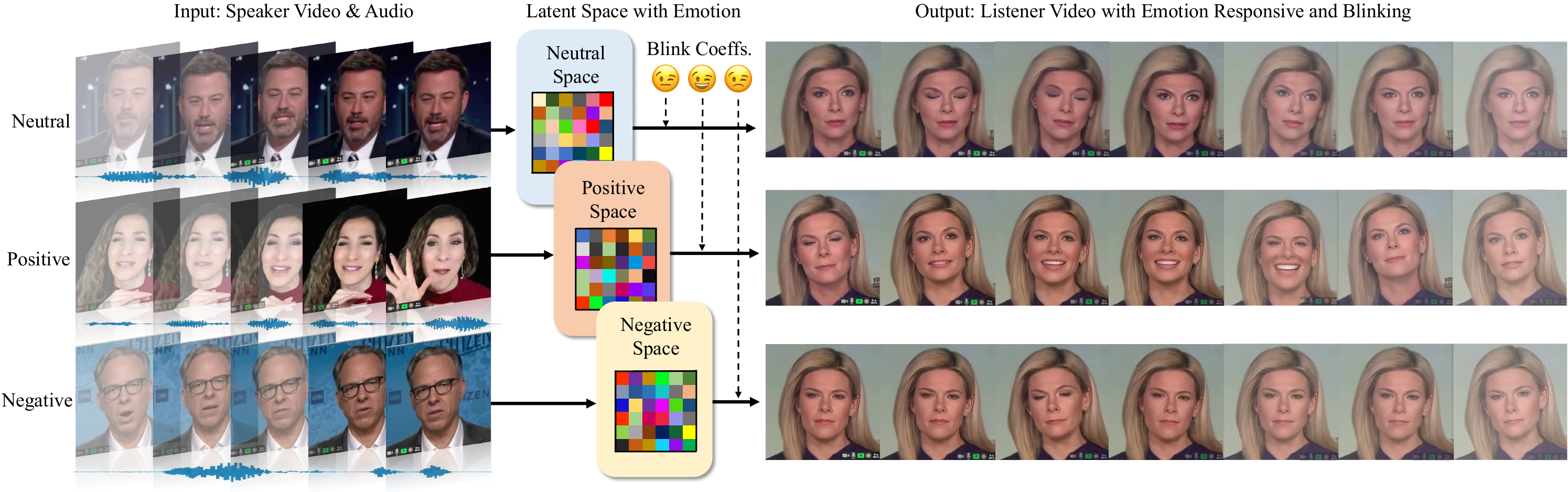}
    \captionof{figure}{Illustration of our method on listener motion synthesis with the ternary emotional value as an example. Given different emotional speakers~(\eg neutral, positive, and negative), our approach generates corresponding listeners under the different emotional latent spaces. 
    }
\label{teaser}
\end{center}%
}]
%%%%%%%%% ABSTRACT
\begin{abstract}
   Listener head generation centers on generating non-verbal behaviors (e.g., smile) of a listener in reference to the information delivered by a speaker. A significant challenge when generating such responses is the non-deterministic nature of fine-grained facial expressions during a conversation, which varies depending on the emotions and attitudes of both the speaker and the listener. To tackle this problem, we propose the Emotional Listener Portrait (ELP), which treats each fine-grained facial motion as a composition of several discrete motion-codewords and explicitly models the probability distribution of the motions under different emotion in conversation. Benefiting from the ``explicit'' and ``discrete'' design, our ELP model can not only automatically generate natural and diverse responses toward a given speaker via sampling from the learned distribution but also generate controllable responses with a predetermined attitude. Under several quantitative metrics, our ELP exhibits significant improvements compared to previous methods. 
\end{abstract}

%%%%%%%%% BODY TEXT
\vspace{-5mm}
\section{Introduction}
\vspace{-1mm}

Listener Head Generation~(LHG) technology aims to synthesize the motion of the listener in response to the speaker. In contrast to speaker head generation~(SHG)~\cite{chen2019hierarchical, song2021tacr, ji2021audio, eskimez2021speech, richard2021meshtalk, suwajanakorn2017synthesizing,chen2020talking, vougioukas2020realistic, song2021talking}, which focuses on generating lip-speech synchronized portrait videos, LHG analyzes the talking semantics of the speaker automatically, without explicit guidance, to synthesize corresponding interactive motions of the listener. As shown in Figure~\ref{teaser}, the listener reacts positively when the speaker shares happy, and vice versa. LHG can be employed in many applications, \eg human-computer interaction~\cite{yu2017talking, liu2022audio,zhang2021flow}, virtual reality~\cite{jonell2020let,li2021ai}, metaverse~\cite{chen2021high, cerekovic2016rapport, song2021fsft} and media forensics~\cite{songadaptive, rossler2018faceforensics, song2022face, he2021forgerynet}~\etc. \par

The distinct nature of LHG, which necessitates a comprehensive modeling of the speaker's motion~\cite{nojavanasghari2018interactive}, presents a significant hurdle in yielding realistic listener head. In the absence of audio-to-mouth matching evaluation, audiences are more inclined to discern subtle changes in facial expressions and head movements. However, the existing methods Responsive Listening Head Generation (RLHG)~\cite{zhou2021responsive} and Learning2Listen~\cite{ng2022learning} have ignored these key components. Specifically, RLHG~\cite{zhou2021responsive} has replicated the regression experience from SHG~\cite{chen2019hierarchical}, which weakens the non-deterministic properties and smoothes the listener motion. Meanwhile, although the motion categories in codebook proposed by Learning2Listen~\cite{ng2022learning} alleviate this problem, the one-dimensional codebook from VQ-VAE~\cite{mirsamadi2017automatic} limits the diversity of motion and emotional representation. Consequently, different emotional states are intricately intertwined within a single codebook, and the generated listener emotion tends to be the biased emotion in the training set. Moreover, neither method can simulate fine-grained facial motion under different emotions, such as minute alterations in the motion surrounding the eyes and movements of the mouth. To explore the solution toward superior LHG results, we focus on the two unresolved hurdles: (1) how to simulate finer-grained listener movements, including the head motion and expression details of the face, and (2) how to explicitly model emotions in the discrete space.\par

In this paper, we propose a novel method called \textbf{E}motional \textbf{L}istener \textbf{P}ortrait (\textbf{ELP}) for vivid listener head video generation. The visual and audio information from the speaker is combined together for the listener motion synthesis (the right part of Figure~\ref{teaser}). (1) To overcome the limitation imposed by one codeword search, we have expanded the classification dimensions to facilitate the mapping of listener motion onto a higher-dimensional discrete space. The fine-grained listener movements correspond in high-dimensional discrete space, which offers greater capacity for the precise depiction of the listener's facial expression and head pose than a single codeword. (2) Despite the expanded latent space on the codeword, explicit emotional representation remains unattainable. As such, building upon the increased space, we leverage emotion priors to split and rearrange the discrete space. More specifically, different emotions are rearranged into corresponding spaces, with the distance between these spaces being determined by the value range of codeword, shown in the middle of Figure~\ref{teaser} (it takes the ternary emotion as an example). The listener features (blink, facial and head motion) from the spaces with emotion are decoded into different emotional listeners, shown in the right part of Figure~\ref{teaser}. It indicates that the distance between listeners with different emotions is widened, such as listeners in positive respone by smiling, while listeners in negative tend to frown. \par 

There are two modules in the \textbf{ELP}, the Adaptive Space Encoder (\textbf{ASE}) and Mesh-to-Video Renderer. In \textbf{ASE}, the discrete latent space obtained by the one-hot vectors argument maximum is concatenated and weighted according to the position with the prior emotion. Employing this approach leads to a further enlarge in the probability distribution distance, and the listener motion coefficients are learned from that. The Mesh-to-Video Renderer renders the photorealistic face from the mesh corresponding to the predicted coefficients with only a single portrait image of the listener. \par

We demonstrate the ability of our method through quantitative and qualitative experiments on two popular conversation portraits datasets, the ViCo~\cite{zhou2021responsive} and the large-scale in the wild conversation videos collected by the Learning2Listen~\cite{ng2022learning}. Our contributions are summarized: \par
\begin{itemize}
\setlength{\itemsep}{0pt}
\setlength{\parsep}{0pt}
\setlength{\parskip}{0pt}
\item[--] We propose a novel framework called \textbf{ELP} for emotional listener head generation in dynamic conversion, which can improve the fidelity of the fine-grained generated listener with facial expression, head pose and blink \etc.  
\item[--] We introduce the Adaptive Space Encoder (\textbf{ASE}) to rearrange the latent space based on emotional priors to obtain more explicit emotional representation.
\item[--] Extensive experiments demonstrate that our method outperforms most existing methods in quantitative and qualitative results. 
\end{itemize}

\section{Related Work}
\label{sec:related}

\begin{figure*}
\begin{center}
\includegraphics[width=1.0\linewidth]{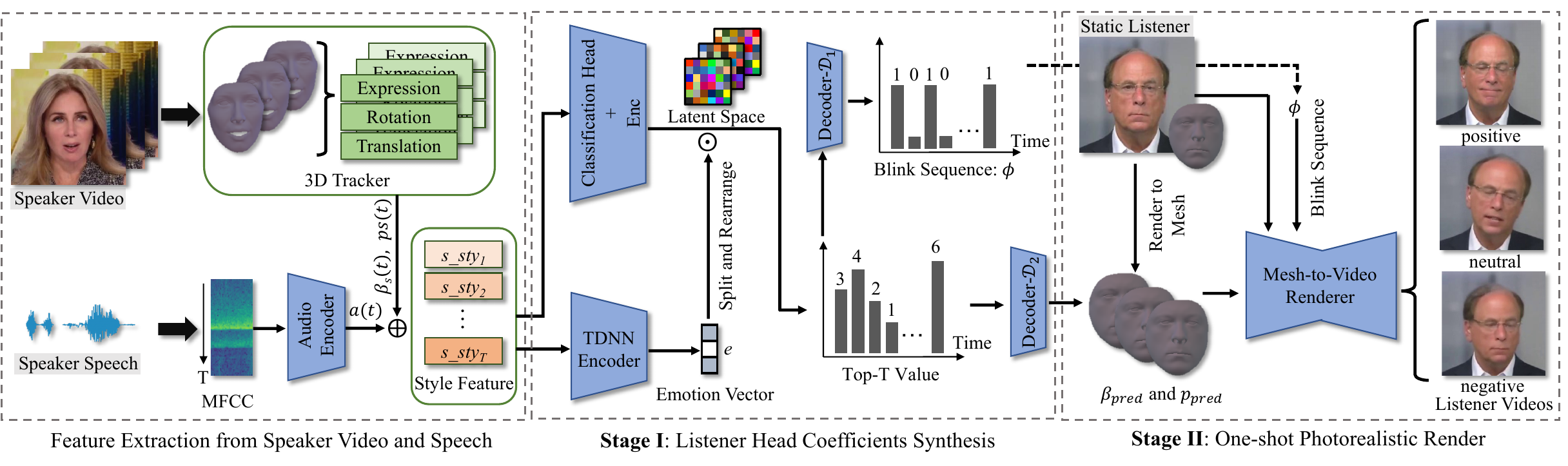}
\end{center}
\vspace{-0.3cm}
  \caption{The overview of \textbf{ELP}. We take the static listener face, speaker video and the corresponding speaker speech as input. In the \textbf{stage I}, the Adaptive Space Encoder maps the discretized features combined with emotion to the the listener motion parameters, then Mesh-to-Video Renderer generates the emotional listener videos from the obtained parameters in the \textbf{stage II}. 
}
\label{fig_overview}
\end{figure*}

\noindent \textbf{Blink-Controlled Facial Animation.}
There are some existing methods~\cite{zhang2021facial, sinha2022emotion} for blink control on facial animation. FACIAL~\cite{zhang2021facial} first employs an eye attention map to locate the eye region, then uses constraints between the rendered pixels and the synthetic result to realize the realistic eye blinks. Sinha~\etal~\cite{sinha2022emotion} adopt the random one-hot encoding for blink control. They use blink one-hot encoding of input into landmark generator. The landmark then guides the generation of the face images. However, these are specifically designed for SHG. Regarding LHG, the methods used to predict blinks through speech are unsuitable as they rely on regression techniques that ultimately lead to a low blink frequency. That can lead to artifacts if the listener keeps their blink frequency extremely strange during a conversation. \par

\noindent \textbf{Audio-Visual Emotion Representation.} There are numerous approaches for extracting emotions from speaker videos. These include Automatic Speech Emotion Recognition~\cite{mirsamadi2017automatic}, Advanced LSTM~\cite{tao2018advanced}, and Cross-Reconstructed Emotion Disentanglement~\cite{ji2021audio}, which aim to decode emotions from speech alone. However, relying solely on speech feature extraction may overlook the visual modality, resulting in insufficient accuracy of the extracted features~\cite{owens2018audio}. Schoneveld~\etal~\cite{schoneveld2021leveraging} employed a joint model of facial and speech features to extract more representative emotional features. \par

\noindent \textbf{Face motion simulation via speech.} It include two parts of work: speaker head generation (SHG) and listener head generation (LHG). Over the past few years, SHG has emerged as a burgeoning field. Chung~\etal~\cite{chung2020seeing} and Shi~\etal~\cite{shi2022learning} extract speech features to simulate the corresponding face or mouth movement. Wu~\etal~\cite{wu2021imitating} imitates arbitrary talking style with speaker speech and head motion features, this method of feature design is also introduced into our work. The LHG is different from the SHG, which pays more attention to the feedback of the listener motion to the speaker. Ahuja~\etal~\cite{ahuja2019react} focuses on the non-verbal behaviours generation in human body, Bohus~\etal~\cite{bohus2010facilitating} and Gratch~\etal~\cite{greenwood2017predicting} study synchronized conversations agent motion in dyadic conversation adapt speech. More recently, Ng~\etal~\cite{ng2022learning} regress the discrete listener head motion with VQVAE~\cite{van2017neural} and Geng~\etal~\cite{geng2023affective} retrieve possible videos of listener face with large language model. In our work, we also learn the non-verbal listener motion in the dynamic communication. \par

\section{Method}
\label{sec: method}

\subsection{Overview} 
\label{subsec: Preliminaries}

We propose a two-stage listener portrait synthesis framework (\textbf{ELP}) which synthesizes emotional listener videos with three inputs: one static portrait of the listener (Static Listener), speaker video (Speaker Video) and the corresponding speech (Speaker Speech). The overview pipeline is shown in Figure~\ref{fig_overview}. We formalize the two stages as follows. \par

\noindent \textbf{Stage I: Listener Coefficients Synthesis.} In this stage, given the speaker video and corresponding speech, the facial and head pose motion ($\beta_{s}(t), p_{s}(t)$) are tracked from the video. It aims to generate the response listener facial head pose movement and blink coefficient sequence from the speech and $\beta_{s}(t), p_{s}(t)$. The ground truth facial and head parameters are captured from the monocular video with Gauss-Newton optimization~\cite{thies2016face2face}, they are resulting in $\beta \in \mathbb{R}^{100T}$ (facial expression), $p \in \mathbb{R}^{6T}$ (head rotation). And the blink coefficients sequence is a binary list on time series ($\phi \in \mathbb{R}^{1T}$). $T$ is the length of video. \par

\noindent \textbf{Stage II: One-shot Photorealistic Render.} In this second stage, our objective is to generate dynamic and photorealistic videos of the listener by utilizing the predicted results from the first stage and a single static portrait image of the listener. \par 

To streamline the problem, we will introduce parameter definitions within our pipeline in Section~\ref{Parameter Definitions}, and go through the details of the two stages in Section~\ref{Listener Head Coefficients Synthesis} and Section~\ref{One-shot Photorealistic Render}, respectively. \par

\subsection{Parameter Definitions} 
\label{Parameter Definitions}

To begin this section, we define the initialization input cross-modal data: the MFCC feature~\cite{logan2000mel} from speech and the facial/head motion parameters $\beta_s (t), p_s (t)$. \par 

\noindent \textbf{Speaker Style Features.} Given the interplay between the speaker and listener, it is necessary to study the stylistic features inherent in the speaker. We expect to employ these features in order to retrieve the emotional underpinnings of the conversation, as well as the listener's response movements. Therefore, we define a computational rule, which extracts speaker style-related features: 
\begin{equation}
s\_sty(t) = a(t) \oplus \sigma(\beta(t)) \oplus \sigma(\frac{\partial \beta(t)}{\partial t}) \oplus \sigma(\frac{\partial p(t)}{\partial t}),
\label{form_xxx}
\end{equation}
the $a(t)$ is the feature of MFCC via Audio Encoder, the $\oplus$ is the concatenation, the $\sigma(\cdot)$ is the standard deviation. The $\beta(t), \sigma(\frac{\partial \beta(t)}{\partial t}), \sigma(\frac{\partial p(t)}{\partial t})$ represent the fluctuation of the speaker's movements relative to time, which can reflect the speaker's visual movement style~\cite{wu2021imitating}. For speech features, some methods~\cite{cudeiro2019capture, thies2020neural} attempt to extract semantic features from pretrained speech-to-text models such as DeepSpeech~\cite{hannun2014deep}. However, it is not applicable to our method, as speech emotion and intonation within is important for listener, but are not included in purely semantic feature. \par

\noindent \textbf{Emotion Vector.} The emotion vector $e$ is a one-hot embedding, which is used to determine which registered emotion codebooks are selected. \par

\noindent \textbf{Blink Coefficients.} We leverage the geometric ratio of eye landmarks to represent eyelid movement. When the ratio~\cite{cech2016real} exceeds a predefined threshold value, the current frame is classified as exhibiting blinking behavior, which is represented by the binary value $1$, otherwise $0$. \par

\subsection{Listener Head Coefficients Synthesis.} 
\label{Listener Head Coefficients Synthesis}

In this stage, we present the Adaptive Space Encoder (ASE) for generating emotional listener coefficients. Specifically, the ASE takes the $s\_sty_{1:T}$ as input. We consider $s\_sty$ to be a cross-modal feature from the speaker, containing various information required by the listener, including emotional value, utterance semantics, and response guidance. The latent speech feature $a(t)$ are obtained from the backbone of ResNet-50 and Dropout from the input MFCC feature, shown in the right part of Figure~\ref{fig_overview}. \par

\begin{figure}
\begin{center}
\includegraphics[width=1.0\linewidth] {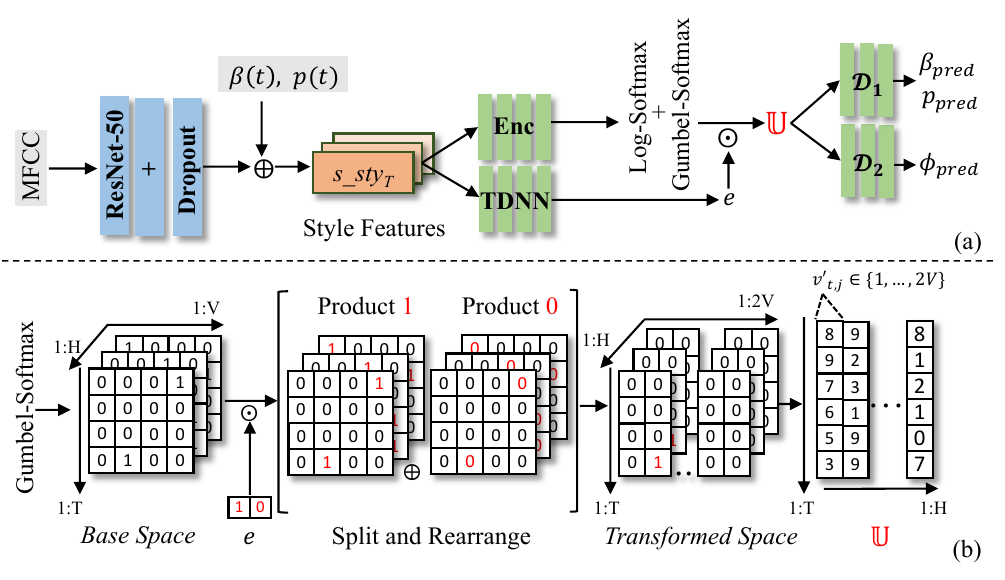}
\end{center}
\vspace{-0.2cm}
  \caption{The structure and details of Adaptive Space Encoder. (a) The Adaptive Space Encoder takes speaker speech MFCC~\cite{logan2000mel} feature, speaker facial ($\beta(t)$) and head motion ($p(t)$) coefficients as input, and outputs the listener's facial and head motion. (b) The details of split and rearrangement for latent space $\mathbb{U}$ (N=2 as an example), the \textit{Base Space} is weighted and concatenated ($\oplus$) by the element values of $e$. 
}
\label{fig_encoder}
\end{figure}

The ASE is shown in the middle of Figure~\ref{fig_overview} and Figure~\ref{fig_encoder}(a), which reveals that ASE is composed of two encoders and decoders. We assume that in a clip of input speaker video ($2$ seconds), the emotion is constant. For the emotion classification branch, the TDNN~\cite{peddinti2015time} Encoder takes a series of $s\_sty_{1:T} \in \mathbb{R}^{25T \times D_s}$ (T is length of video with $25$ framerate, $D_s$ is the dimension of each $s\_sty(t)$) as input and encodes them into the predicted emotion vector $e_{pred}$. For the motion prediction branch, the ASE model leverages a multi-classification head mechanism to encode the continuous-valued stylized features to a discrete latent space, which is a classification probability distribution and then the predicted motions are sampled from this distribution. Through this mechanism, we obtain a motion space with discrete manifold. We achieve discretization of this categorical representation with gumbel-softmax~\cite{jang2016categorical,richard2021meshtalk}, which can be formalized by: 
\begin{equation}
\rm{v_{t; h; 1} = [Gumbel\text{-}Softmax(\textbf{enc}(s\_sty)_{t, h, 1:V})]_{1:H; 1}},
\label{form_gumbel}
\end{equation}
where the $\rm{\textbf{enc}(s\_sty)_{t, h, 1:V}}$ means the multi-modality feature $s\_sty$ encoded to $\rm{T \times H \times V}$-dimensional latent, which is still in continuous-value space. Then, the operation of Gumbel-Softmax~\cite{jang2016categorical} takes the $\rm{\textbf{enc}(s\_sty)_{t, h, 1:V}}$ into the maximum probability in V dimension, which represents the corresponding discrete motion code in the space be mapped by Equation~\ref{form_gumbel}, shown in Figure~\ref{fig_encoder}(b). The $\rm{v}_{t;h;1}$ means each codeword value in the discrete space, the H is the number of latent classification heads and V is the number of categories. For the sake of simplicity, we refer to this space as the \textit{Base Space} throughout the remainder of this paper, the \textit{Base Space} is composed of one-hot vectors. However, emotions are still implicit in the \textit{Base Space}. To capture these emotional responses explicitly, it is necessary to distinguish different emotion spaces. \par

Specifically, we split and rearrange the \textit{Base Space} based on the emotion prior. The emotion vector $e$ perform dot production with \textit{Base Space}, and the results are concatenated. The $e \in \mathbb{R}^{1 \times \rm{N}}$ is a one-hot vector, N is the number of emotion types. Then the \textit{Base Space} is expanded to $\rm{T \times H \times NV}$ from $\rm{T \times H \times V}$. We call the expanded result the \textit{Transformed Space}. Finally, by computing the argument maximum on \textit{Transformed Space}, we obtain the final discrete latent space $\mathbb{U}$. The codeword value $\rm{v'}$ in the $\mathbb{U}$ is in the range of $\{\rm{v'_{1:T,1:H} | v'_{i,j} \in [1,2,...,NV]}\}$. It should be noted that the values within different emotional intervals are unique and non-overlapping. As an example, let us consider the two different emotion values for $e$. The $\mathbb{U}$ corresponding to the first emotions will only take values in the range of $[1,...,V]$, and the other in $[V+1,...,2V]$. And it can be found that our method obtains $\mathbb{U}$ space without additional dimensional overhead after splitting and rearrangement. Then, two different decoders $\mathcal{D}_1$, $\mathcal{D}_2$ act on the $\mathbb{U}$ to obtain the facial/head movements coefficients and blink sequences. \par

For the training of ASE model, we adopt the parameter series $\beta' (t), p'(t)$ reconstructed from the monocular listener video as ground truth, and calculate $s\_sty(t)$ frame by frame. Afterwards, we feed $s\_sty(t)$ to the ASE model, yielding the predicted listener parameter $\beta _{pred}(t)$, $p_{pred}(t)$ from $\mathcal{D}_1$ and blink sequence $\phi_{pred}(t)$ from $\mathcal{D}_2$. Based on the predicted $\beta _{pred}(t)$ and $p_{pred}(t)$, we apply the $L_2$ loss as follows:
\begin{equation}
\mathcal{L}_{L_{2}} = \sum_{t=1}^{T}||\beta _{pred}(t) - \beta'(t) ||_2 + ||p_{pred}(t) - p'(t) ||_2. 
\label{form_1}
\end{equation}
We take the blink sequence as a decision at each time instance $t$, and employ the cross-entropy loss for binary classification, the $\phi(t)$ is the ground truth blink state:
\begin{equation}
\mathcal{L}_{\rm{CE}_1} = -\sum_{t=1}^{T} \phi(t) log \phi_{pred}(t)+[1-\phi(t)]log [1-\phi_{pred}(t)], 
\label{form_2}
\end{equation}
we introduce regularization loss to suppress noises and encourage a more concentrated density distribution (a blinking action comprises several consecutive 1 value): 
\begin{equation}
\mathcal{L}_{\rm{reg}} = \sum_{t=2}^{T} || \phi_{pred}(t) - \phi_{pred}(t-1) ||_1.
\label{form_3}
\end{equation}
Additionally, for emotional constraints, TDNN encoder encodes $s\_sty$ into vector $e'$, we also apply cross entropy on it, the $e$ is the groundtruth one-hot emotion vector: 
\begin{equation}
\mathcal{L}_{\rm{CE}_2} = -\sum_{i=1}^{N} e_i log e'_i+[1-e_i]log [1-e'_i],
\label{form_4}
\end{equation}
which is used for $N$ classification of emotions, $N$ is the type of emotion. The final loss function is then defined as:
\begin{equation}
\mathcal{L}= \mathcal{L}_{L_{2}} + \lambda_1 \mathcal{L}_{\rm{CE}_1} + \lambda_2 \mathcal{L}_{\rm{CE}_2} + \lambda_3 \mathcal{L}_{\rm{reg}},
\label{form_5}
\end{equation}
the $\lambda_1, \lambda_2$ and $\lambda_3$ are three weight to balance these terms.

\subsection{One-shot Photorealistic Render} 
\label{One-shot Photorealistic Render}
To improve the visual generalization in our method, we also train a one-shot Mesh-to-Video translation network. Usually, it is very difficult to collect long-term person-specific portrait videos to train high-quality person-specific video, so we tend to use one-shot rendering method, which is resource free. Inspired by Siarohin~\etal~\cite{siarohin2019first}, we first train the motion capture module~\cite{siarohin2019first} on face mesh videos, and then use the face mesh videos to drive the motion of one-shot portrait image to synthesis the photorealistic video, we call it as Mesh-to-Video Renderer, shown in the right part in Figure~\ref{fig_overview}. To control the blinking, we obtain the length $L$ of each eye blink motion group on $\phi$ (each consecutive $1$ is counted as one group), and the expression blendshape $\beta_1, \beta_2$ at the beginning and end of the motion group (the eye closure blendshape is marked as $\beta_c$). We interpolate $\beta_1 \rightarrow \beta_c$ and $\beta_c\rightarrow \beta_2$ respectively on the interval of $L/2$. The blendshape for eye blink is linearly weighted, simulating the eyelid position at each timestamp for physical blink or emotional events around eyes. \par

\begin{table*}[t]
\footnotesize
\vspace{-0.25cm}
\begin{center}
\setlength{\tabcolsep}{1.55mm}{
\begin{tabular}{cccccccccccccccc}
\hlinew{1.15pt}
\multirow{3}{*}{Methods} &\multicolumn{6}{c}{Facial \textit{Coeff.} ($\beta$)} & & \multicolumn{6}{c}{Head \textit{Coeff.}($p$)} & & \multicolumn{1}{c}{Blink ($\phi$)}\\
\cline{2-7} \cline{9-14} \cline{16-16}
&FD$\downarrow$ &V-D &SI-D &RPCC$\downarrow$ &WTLCC &STS$\downarrow$ &      &FD $\downarrow$ &V-D &SI-D &RPCC$\downarrow$ &WTLCC &STS$\downarrow$ & & WTLCC$\uparrow$\\
~&($\times 10^3$)& & &  & ($\rightarrow$GT)  & ($\times 10^3$) & &($\times 10^3$)& & & & ($\rightarrow$GT) &($\times 10^3$) & &  \\
\hline
NN-motion &12.17 &2.88 & 2.70 &0.16  &0.092 &8.89 &      &6.65&1.79&1.16&0.20&0.083&5.59& &0.12\\
NN-audio &19.05 &3.70 &2.99 &0.18 &0.077 &11.36 &     &14.05& 2.51 &2.54&0.15&0.123 &7.10 &  &0.13\\
Random &56.21 &4.57 & 3.92 &0.28 &0.047 &40.26    &     &35.10 &2.54 & 2.29 & 0.15 &0.043 &25.08&  &0.11\\
DLS-Random &38.10 &4.01 &3.47 &0.21  &0.075&22.51&      &26.17 &3.99 &1.33 & 0.12 &0.105 &15.50&  &0.11\\
RLHG~\cite{zhou2021responsive}&150.0$^{*}$&3.72&0.41 &0.20 & 0.011 &29.00 &   &143.1$^{*}$&0.28 &0.24 &0.16 & 0.010 &25.64&  &0.04\\
PCH$^{**}$~\cite{huang2022perceptual}&18.10 & 1.14 & 0.37 &0.18 &0.011 & 27.22 &    & 20.35  &0.24 &0.22&0.10 &0.003 &17.44& 
 &$1e^{-2}$\\
L2L~\cite{ng2022learning}&4.20 & 3.02 &2.79 &0.11 &0.094 & 5.44  &         &1.93 &0.97 &2.45 &0.04 &0.017 &3.38 & &$1e^{-2}$\\
\rowcolor{gtgray} GT &-- & 4.89 & 4.27 & -- & 0.177 & -- &        &--&1.90 & 1.96 &--&0.192 &--&  &--\\
\rowcolor{mygray} \textbf{ELP} &\textbf{1.14} &5.26 &4.00  &\textbf{0.02} &\textbf{0.166}  &\textbf{2.93} &   & \textbf{0.60} &2.04 &1.55 &\textbf{0.01} &\textbf{0.199} &\textbf{1.34} &  & \textbf{0.39}\\
\hline
NN-motion &15.80 &2.14  & 2.37 & 0.12 &0.155 & 9.33  &        & 7.97 & 1.55 & 1.92 &0.13 &0.061  &7.94 &  &0.12\\
NN-audio &22.03 &1.87 & 1.61  &0.09 &0.114 &8.90  &    &13.37 &1.39 &1.20 &0.11 &0.077 &7.15 & &0.09\\
Random &47.02 & 2.26 & 2.51 &0.58 & 0.044 & 20.41   &       &24.70 &1.84 &1.91 &0.19 &0.075 &13.80 & &0.11 \\
%%%%%%%%%%%%%%%%%%
DLS-Random& 33.61 &2.23 &2.14 & 0.12 &0.071 &8.04 &            &19.97 &1.37 &1.54 &0.09 &0.088 &6.29 &  &0.13 \\
RLHG~\cite{zhou2021responsive}&21.28 & 0.59 & 1.06 & 0.39 & 0.070 & 14.11&        &18.54 &0.14 & 1.10 &0.292 &0.026 &12.31  &  & 0.02\\
L2L~\cite{ng2022learning} & 3.55$^*$ & 2.01$^*$ & 2.48$^*$ & 0.02$^*$ & 0.130 & 7.39 &    &0.81$^*$ &0.62$^*$ &1.82$^*$ &\textbf{0.00}$^*$ & 0.004 &6.01 & &$1e^{-2}$\\
\rowcolor{gtgray} GT &--  & 2.47 & 2.20 & -- & 0.202 & --    &       &--&0.55 &1.54 & -- &0.127 & -- &  & -- \\
\rowcolor{mygray} \textbf{ELP} &\textbf{1.37} & 2.70 &2.15 &\textbf{0.014} &\textbf{0.182} &\textbf{4.49} &     &\textbf{0.36} &0.59 &1.60 &0.077 & \textbf{0.130} &\textbf{2.51}& &\textbf{0.42}\\
\hlinew{1.15pt}
\end{tabular}}
\end{center}
\caption{Quantitative results on two different datasets. The above is the performance evaluation on the ViCo dataset~\cite{zhou2021responsive}, whereas the bottom is on the L2L dataset~\cite{ng2022learning}. The $\downarrow$ indicates lower is better, and $\rightarrow$GT means closer to GT is better. The $*$ indicates that we directly follow the office report results and $**$ means we reproduce the PCH~\cite{huang2022perceptual} on our system (no source code is provided). The best performances are highlighted in bold.}
\label{table_all_compare}
\end{table*}

\section{Experiments}
\subsection{Experimental Settings} 
\label{Experimental Settings}

\noindent \textbf{Datasets.} We evaluate our method on two of the most popular conversation portrait datasets, the ViCo~\cite{zhou2021responsive} dataset and the dataset proposed by Learning2Listen~\cite{ng2022learning}. The ViCo dataset contains rich samples of 483 video clips with 50 different listener identity and three emotion annotations (we set $N=3$ for emotion for this dataset). The Learning2Listen~\cite{ng2022learning} is a 72 hours versus 95 minutes dataset collected in the wild, which comes from Youtube with six identities but no emotion annotations. We adopt the pretrained speech emotion analyzer model~\cite{Speech-Emotion-Analyzer} to extract emotions. We set the [``happy''] to positive and [``calm'', ``fearful'', ``sad'', ``angry''] to unpositive (we set $N=2$). To maintain the balance in this dataset, we assign more labels to unpositive emotions (the positive videos more than other emotions in this dataset). The one-shot photorealistic renderer is trained on the TalkingHead-1KH~\cite{wang2021one} datasets with $256 \times 256$ resolution. 

\noindent \textbf{Network Architectures.} We apply the first 16 layers of ResNet-50 as the backbone to encode the input MFCC from $\rm{29T}$-dimension to $\rm{128T}$-dimension. The TDNN Encoder is composed of five hidden layer TDNN and three-layer MLP, which is used for the classification of $N$-dimension emotional one-hot vector from the $\rm{334T}$-dimension ($\rm{128T}$ for a(t), $\rm{100T}$ for $\sigma (\beta(t))$ and $\sigma(\frac{\partial \beta(t)}{\partial t})$, $\rm{6T}$ for $\sigma(\frac{\partial p(t)}{\partial t})$) speaker style feature. The Classification Head Encoder consists of 8 layers conv1d and 3 layers of LSTM with MLP, it takes the style feature as input, the dimension of output is $\rm{T \times H \times 3C}$. The decoder $\mathcal{D}_1$ has two conv2d layers and the decoder $\mathcal{D}_2$ has four layers conv2d with three layers LSTM, their role is to recover the motion from latent space. The structure of one-shot photorealistic renderer (Mesh-to-Video Render) is from the First Order Motion Model~\cite{siarohin2019first}. 

\noindent \textbf{Implementation details.} For the convenience of training, we randomly clip the input to 2 second video (50 frames), and clip the output corresponding to 50 frames. The $\lambda_1, \lambda_2$ and $\lambda_3$ are $5, 5, 0.01$. We set V and H to $64$, $128$ for the latent space. When optimizing, we adopt the AdamW optimizer~\cite{loshchilov2017decoupled} to train the ASE model with the initial learning rate of $1 \times 10^{-3}$. We train $10000$ iterations with a batch size of $32$ samples for ViCo~\cite{zhou2021responsive} dataset and $50000$ iterations with $128$ batch size for Learning2Listen dataset~\cite{ng2022learning}. \par

\noindent \textbf{Baselines.} We compare with the state-of-the-art methods and the hand-craft methods. For the listener's head generation, the hand-craft baseline is usually strong enough. 
\begin{itemize}
\setlength{\itemsep}{0pt}
\setlength{\parsep}{0pt}
\setlength{\parskip}{0pt}
\item[--] Learning2Listen (L2L)~\cite{ng2022learning}: It maps the motion patterns to realistic movements through the VQ-VAE~\cite{van2017neural} without the consideration of emotions and blinks. 
\item[--] Responsive Listening Head Generation (RLHG)~\cite{zhou2021responsive}: It regresses the listener motion from the speaker and audio. We use the official code for fairness.
\item[--] Perceptual Conversational Head (PCH)~\cite{huang2022perceptual}: It wins the $1^{st}$ in the ViCo challenge. Since only part of the code has been released, we reproduce it based on RLHG~\cite{zhou2021responsive}. 
\item[--] NN-motion/NN-audio: For arbitrary input speaker motion or audio input, we find its nearest neighbor from the training set and use its corresponding listener motion as output. We follow it from the L2L~\cite{ng2022learning}.
\item[--] Random: We random select the facial and head motion parameters in the training data, and inject random small perturbations into the normal distribution.
\item[--] Discrete-Latent-Space Random (DLS-Random): Randomly generate $\rm{v}'_{t;1:H}$ value, and use the pretrained $\mathcal{D}_1, \mathcal{D}_2$ to generate motion sequences. 
\end{itemize}

\noindent \textbf{Metrics.} We choose the following metrics to evaluate the generated facial/head motion coefficients. 
\begin{itemize}
\setlength{\itemsep}{0pt}
\setlength{\parsep}{0pt}
\setlength{\parskip}{0pt}
\item[--] Frechet Distance (FD)~\cite{heusel2017gans}: The $L_1$ distance to measure the difference between the generated facial/head motion and the ground truth. 
\item[--] Variation for Diversity (V-D): Proposed by the L2L~\cite{ng2022learning}, the variance of the facial/head motion on time series. 
\item[--] SI for Diversity (SI-D)~\cite{zhang2020generating}: From the L2L~\cite{ng2022learning}, it measures diverseness of predictions with k-means to facial/head motion, we report the average entropy (Shannon index). 
\item[--] Residual Pearson Correlation Coefficient (RPCC): The Pearson Correlation Coefficient (PCC)~\cite{messinger2009automated, riehle2017quantifying} measure facial/head motion frame-by-frame, it is for listener covaries with the input speakers. We calculate the $L_1$ distance between generated PCC and PCC of ground truth. 
\item[--] Windowed Time Lagged Cross Correlation (WTLCC)~\cite{boker2002windowed}: It is the correlation between the generated motion and the input speaker's motion in the set time window. We calculate the $L_1$ distance between generated TLCC and ground truth with the window size as 4 seconds.
\item[--] Short Time Series distance (STS)~\cite{moller2003fuzzy}: The STS can measure the similarity on sampled data. We calculate the STS between generated listener motion and ground truth listener motion. 
\end{itemize}
We report head and facial metrics separately. For blink sequence evaluation, we adopt the WTLCC, which has a good performance on handling the noise by the time shifting, especially for the binary sequences. It is worth noting that for the blink WTLCC we compute the correlation between the generated blink sequence and the real blink sequence. It is different from the WTLCC calculated on the facial and head motion. Please refer to the Appendix for the details about each metric.

\subsection{Comparison Results} 
\label{Comparison Results}

\noindent \textbf{Quantitative Results.} We retrain PCH~\cite{huang2022perceptual} and L2L~\cite{ng2022learning} on the ViCo datasets~\cite{zhou2021responsive}, retrain PCH~\cite{huang2022perceptual} and RLHG~\cite{zhou2021responsive} on L2L~\cite{ng2022learning} datasets. Table~\ref{table_all_compare} shows the quantitative comparison of listener head generation in the first stage. Based on the metrics presented in Table~\ref{table_all_compare}, it is clear that our proposed \textbf{ELP} method outperforms other existing methods by a significant margin. Our method is about $400$ times better than the current state-of-the-art method L2L~\cite{ng2022learning} in terms of facial blink indicators (WTLCC in Blink $\phi$), the L2L~\cite{ng2022learning} struggles to synthesize realistic eye movements, our proposed \textbf{ELP} is capable of generating blinking movements that accurately reflect the conversation. In the realm of measuring motion diversity (V-D, SI-D), L2L~\cite{ng2022learning} rely on the distance from ground truth as a performance metric, but we opt for a more comprehensive perspective. Our expectation is that listeners' movements will showcase a greater range of diversity when the FD and STS metric (feature distance) are kept to a minimum. As demonstrated in Table~\ref{table_all_compare}, it is evident that even when FD and STS are kept to a minimum, our \textbf{ELP} approach remains highly effective in both the V-D and SI-D, which benefits from our latent space split and rearrange method. The RPCC and WTLCC (in Facial and Head \textit{Coeff.}) measures the motion synchronization between the speaker and the listener, and it can be found that our method achieves an effect close to GT (the RPCC of GT is $0$).\par

The weaker performance of regression-based methods~\cite{zhou2021responsive,huang2022perceptual} demonstrates the regression methods are not suitable for the listener head synthesis, due to the inherent randomness involved in listener. This fact is made especially clear when evaluating key indicators such as FD, STS, V-D and SI-D \etc. Despite L2L's~\cite{ng2022learning} impressive performance, it still falls behind our proposed method. This is primarily attributed to the fact that the codebook space of VQ-VAE is equivalent to an encoding head with H=1, while our approach not only considers emotional rearrangement but also utilizes a larger discrete coding space (H=128). The NN-motion/audio and Random/DLS-Random are some hard-to-beat baselines in terms of variance (V-D, SI-D), but they fall short in simulating motion correlation. In other words, their diversity is primarily based on random generation. As mentioned above, our method outperforms the baseline method in both diversity and motion correlation simulation. \par

\begin{figure}[t]
\begin{center}
\includegraphics[width=1\linewidth] {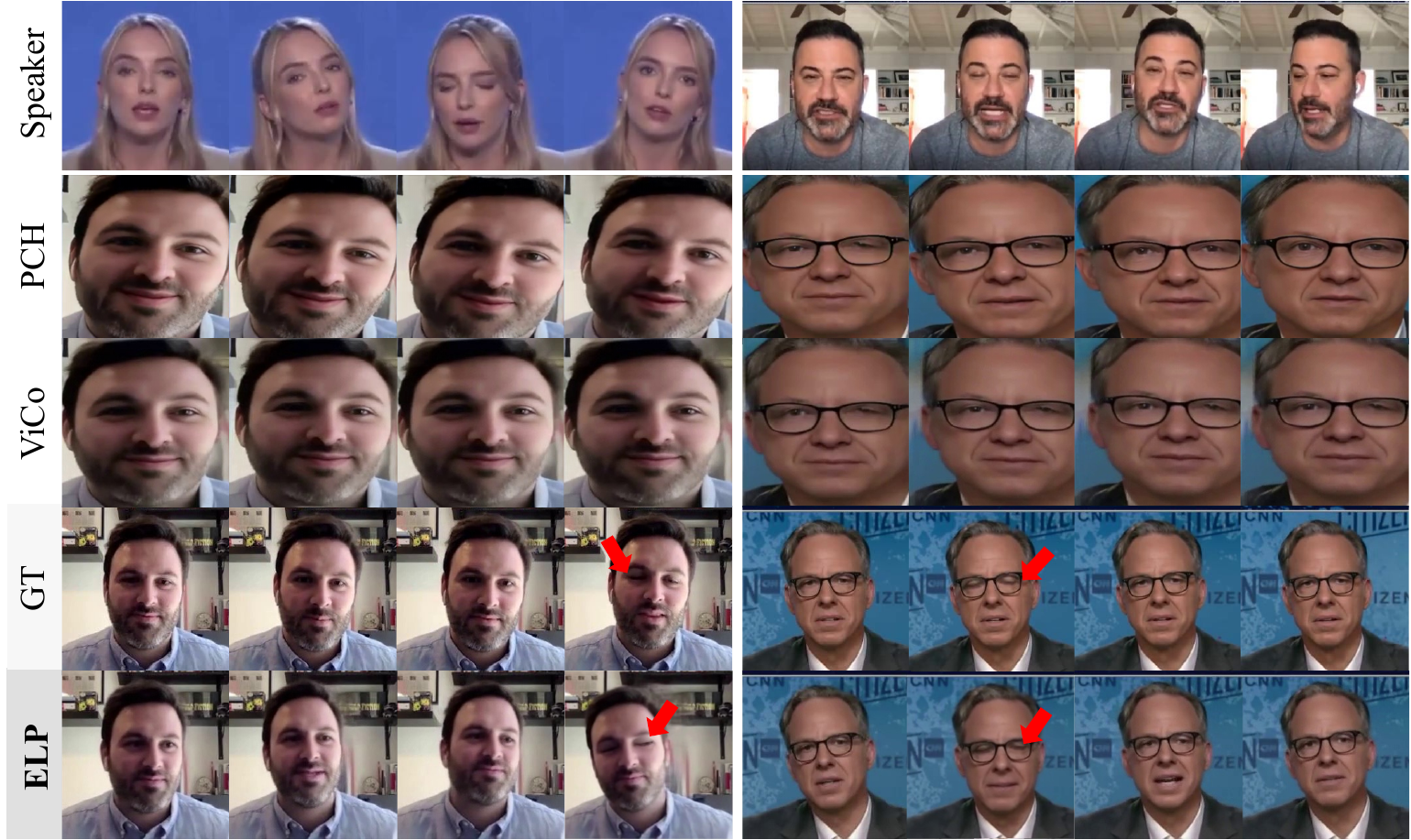}
\end{center}
\vspace{-0.1cm}
  \caption{Comparisons with RLHG~\cite{zhou2021responsive} and PCH~\cite{huang2022perceptual}. Our method synthesizes more diverse facial and head movements, the blinking frames are highlighted with red arrow. The input static listener image from the same one, the speaker videos come the test set of ViCo datasets~\cite{zhou2021responsive}.
}
\label{fig compare}
\end{figure}

\begin{figure}[t]
\begin{center}
\includegraphics[width=1\linewidth] {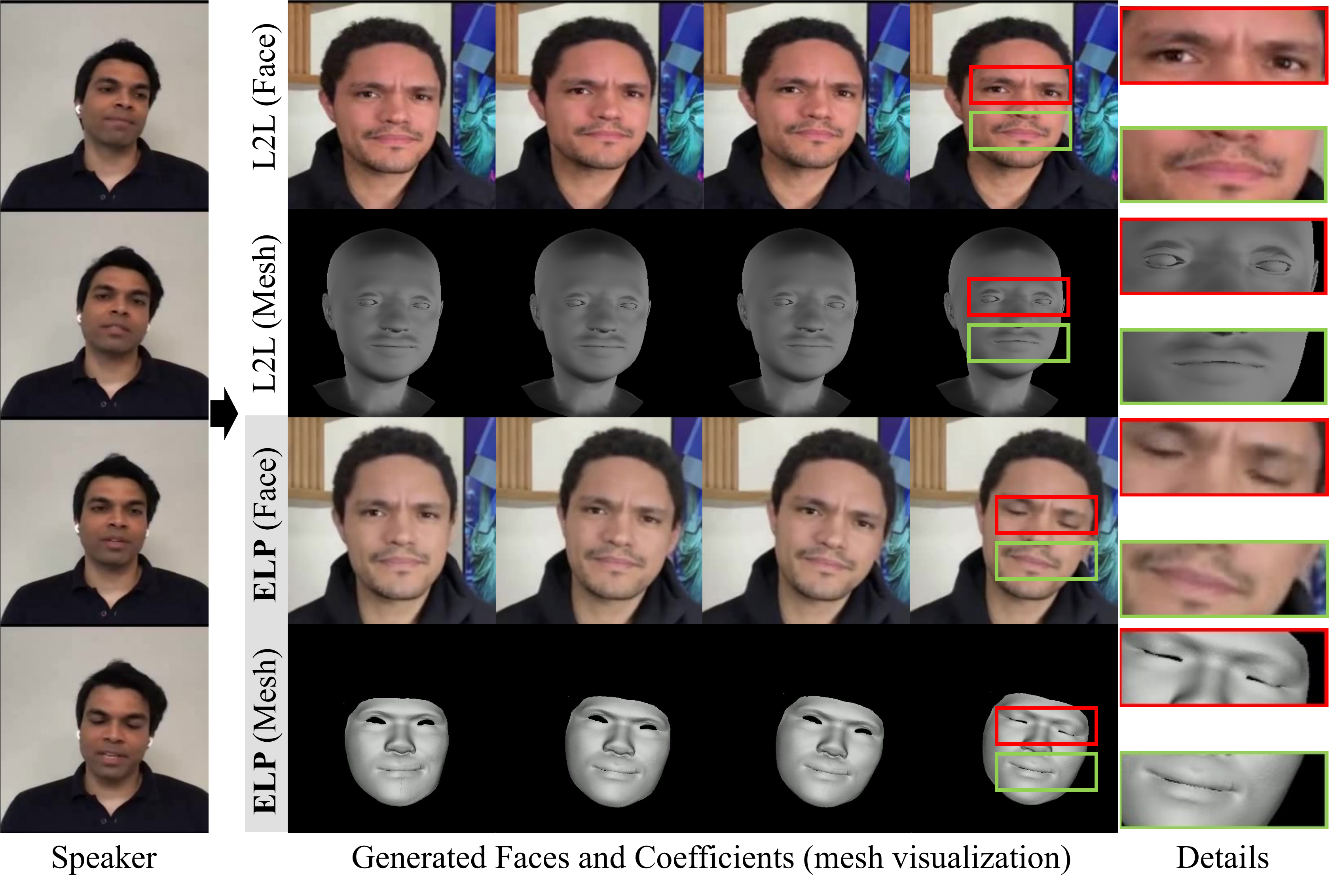}
\end{center}
\vspace{-0.1cm}
  \caption{Comparisons with L2L~\cite{ng2022learning}. The L2L~\cite{ng2022learning} works on the facial and head coefficients level, we visualize the coefficients to mesh for fair comparison. The details of mesh and face are highlighted on the right part.}
\label{fig compare L2L}
\end{figure}

\noindent \textbf{Qualitative Results.} We first compare with the photorealistic results of PCH~\cite{huang2022perceptual} and RLHG~\cite{zhou2021responsive} in Figure~\ref{fig compare}, for fairness, we manually align and compare the results displayed on the office website~\cite{Leaderboard}. The listener head synthesized by RLHG~\cite{zhou2021responsive} and PCH~\cite{huang2022perceptual} exhibit two significant shortcomings compared to ours, 1) the facial movements are limited and lack variation, 2) the videos generated do not incorporate any background information, the \textbf{ELP} overcomes these issues. The visualization results compared with L2L~\cite{ng2022learning} are shown in Figure~\ref{fig compare L2L}, we capture the speaker videos and generate the photorealistic results with Vid2Vid~\cite{wang2018video}, which is a person-specific trained renderer. Since there is no guidance and source code on the settings of renderer in the L2L report~\cite{ng2022learning}, we follow the guidance of Vid2Vid~\cite{wang2018video} office code for training and generation. We also compare the facial expression and head pose details through mesh visualization in Figure~\ref{fig compare L2L}, which excludes the background and facial texture interference. It can be found that the results synthesized by our method have more diversity than L2L~\cite{ng2022learning} in terms of parameters (face mesh) and photorealistic generated listener. \par

\begin{table}[t]
\footnotesize
\vspace{-0.25cm}
\begin{center}
\setlength{\tabcolsep}{0.50mm}{
\begin{tabular}{ccccccccc}
\hlinew{1.15pt}
\multirow{3}{*}{Methods} &\multicolumn{6}{c}{\textit{Coeff.} ($\beta+p$)} & & \multicolumn{1}{c}{Blink ($\phi$)}\\
\cline{2-7} \cline{9-9} %\cline{16-16}
&FD$\downarrow$ &V-D &SI-D &RPCC$\downarrow$ &WTLCC &STS$\downarrow$ & & WTLCC$\uparrow$\\
~&($\times 10^3$)& & & & ($\rightarrow$GT)  & ($\times 10^3$) & &   \\
\hline

NN-motion+$\mathbb{V}$ &18.30 & 3.52  & 3.77 & 0.59 & 0.104 &19.91  &  &0.12\\
NN-motion+$\mathbb{U}$ &\uline{12.82} &4.67  & 3.86 & \uline{0.36} & \uline{0.175} & 14.48  &  &0.12\\
NN-audio+$\mathbb{V}$ &34.49 & 4.13 & 4.24  & 0.51 & 0.128 & \uline{16.67}  & &0.05\\
NN-audio+$\mathbb{U}$ &\uline{33.10} & 6.21 & 5.53  &\uline{0.33} &\uline{0.200} &18.46  & &\uline{0.09}\\
DLS-Rand.+$\mathbb{V}$ &68.02 &6.94 &5.99  &\uline{0.31} &0.120 &39.07 &  &\uline{0.14}\\
DLS-Rand.+$\mathbb{U}$ &\uline{64.27} &8.00 &4.80 &0.33  &\uline{0.180}&\uline{38.01}&  &0.11\\
\rowcolor{gtgray} GT &-- & 6.79 & 6.23 & -- & 0.369 & -- &  &--\\
\rowcolor{mygray} Our+$\mathbb{V}$ &4.27 &3.64 &3.31  & 0.09  & 0.190 &6.12 &  &0.31\\
\rowcolor{mygray} Our+$\mathbb{U}$ &\uline{1.74} &7.31 &5.55  &\uline{0.03}&\uline{0.366} &\uline{4.27}&  & \uline{0.39}\\
\hline
NN-motion+$\mathbb{V}$ &26.43 & 3.77  & 3.91 & 0.26 & 0.207 & \uline{14.33}  &  & \uline{0.15}\\
NN-motion+$\mathbb{U}$ &\uline{23.77} &3.69 & 4.29 & \uline{0.25} & \uline{0.216} & 18.27  &  &0.12\\
NN-audio+$\mathbb{V}$ &49.18 & 5.12 & 4.77 & 0.34 & 0.292 & 27.43 & &0.05\\
NN-audio+$\mathbb{U}$ &\uline{35.40} & 3.26 & 2.81  & \uline{0.20} &\uline{0.191} &\uline{16.05}  & &\uline{0.09}\\
DLS-Rand.+$\mathbb{V}$ &62.38 & 7.83 & 6.06 &0.35  &0.107 &27.04 &  &0.07\\
DLS-Rand.+$\mathbb{U}$ &\uline{53.58} &3.60 &3.68 &\uline{0.21}  &\uline{0.159}&\uline{14.33}&  &\uline{0.11}\\
\rowcolor{gtgray} GT &-- & 3.02 & 3.74 & -- & 0.329 & -- &  &--\\
\rowcolor{mygray} Our+$\mathbb{V}$ &5.12 &2.11  & 1.79 & 0.122  & 0.297 &8.38 &  &0.36\\
\rowcolor{mygray} Our+$\mathbb{U}$ &\uline{1.73} &3.29 & 3.75  &\uline{0.091} &\uline{0.312} &\uline{7.02}&  & \uline{0.42}\\
\hlinew{1.15pt}
\end{tabular}}
\end{center}
\vspace{-0.1cm}
\caption{Ablation study results for latent space $\mathbb{U}$ (with emotional representation) and $\mathbb{V}$ (without emotional representation) on two different datasets. The above is on the ViCo dataset~\cite{zhou2021responsive}, the bottom is on the L2L dataset~\cite{ng2022learning}. The ``DLS-Rand.'' is the baseline method Discrete-Latent-Space Random. The best results in each group are underlined.}
\label{table_all_ab}
\end{table}

\subsection{Ablation Studies}
\label{Ablation Studies}

\begin{figure}[t]
\begin{center}
\includegraphics[width=1\linewidth] {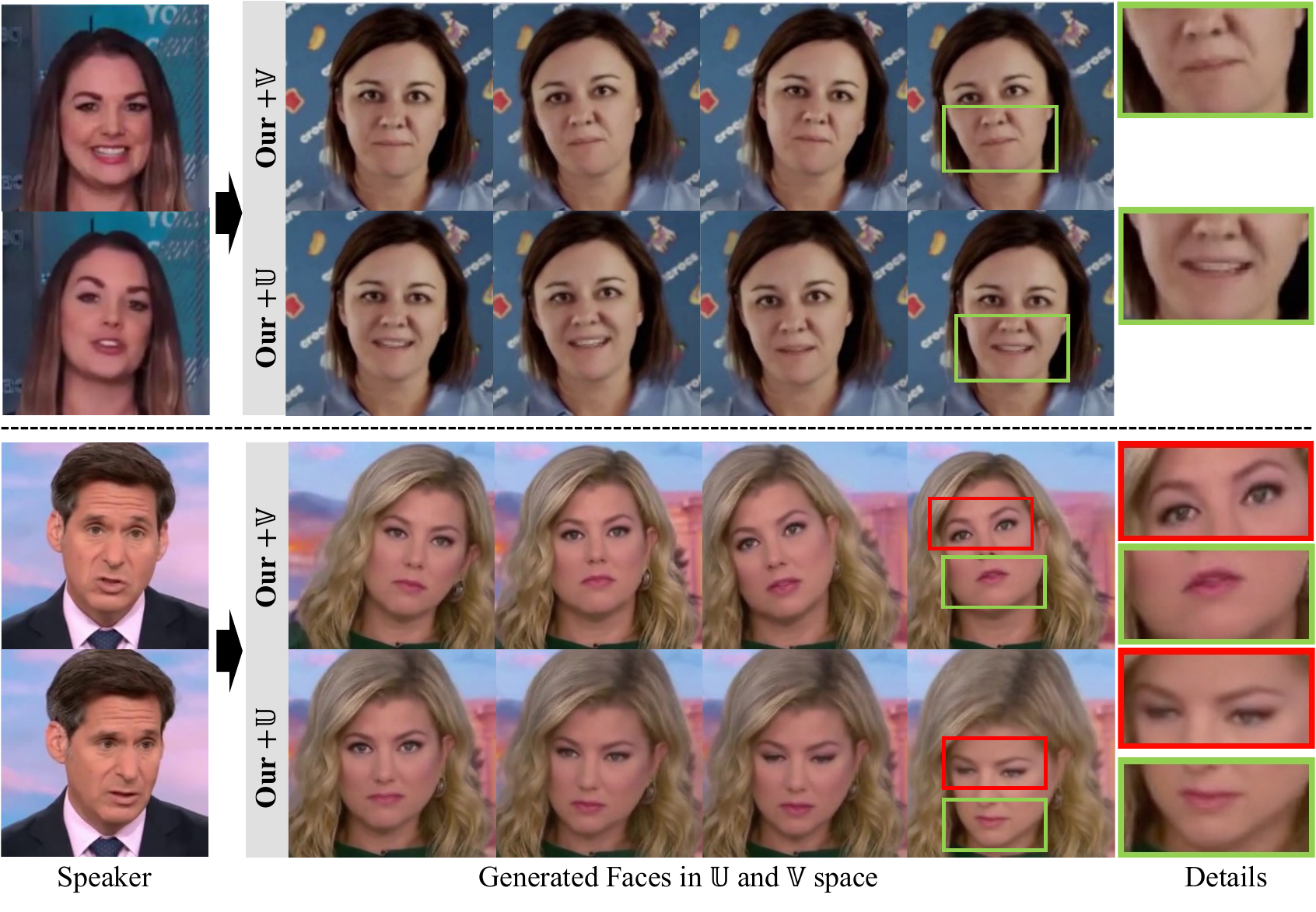}
\end{center}
\vspace{-0.1cm}
  \caption{The visualization results of ablation study on the latent space rearrangement ($\mathbb{U}$ and  $\mathbb{V}$). The upper and lower parts correspond to positive and negative respectively. The details of generated face are on the right part.}
\label{fig compare ab face}
% \vspace{-0.1cm}
\end{figure}

\begin{figure}[tp]
\begin{center}
\includegraphics[width=1\linewidth] {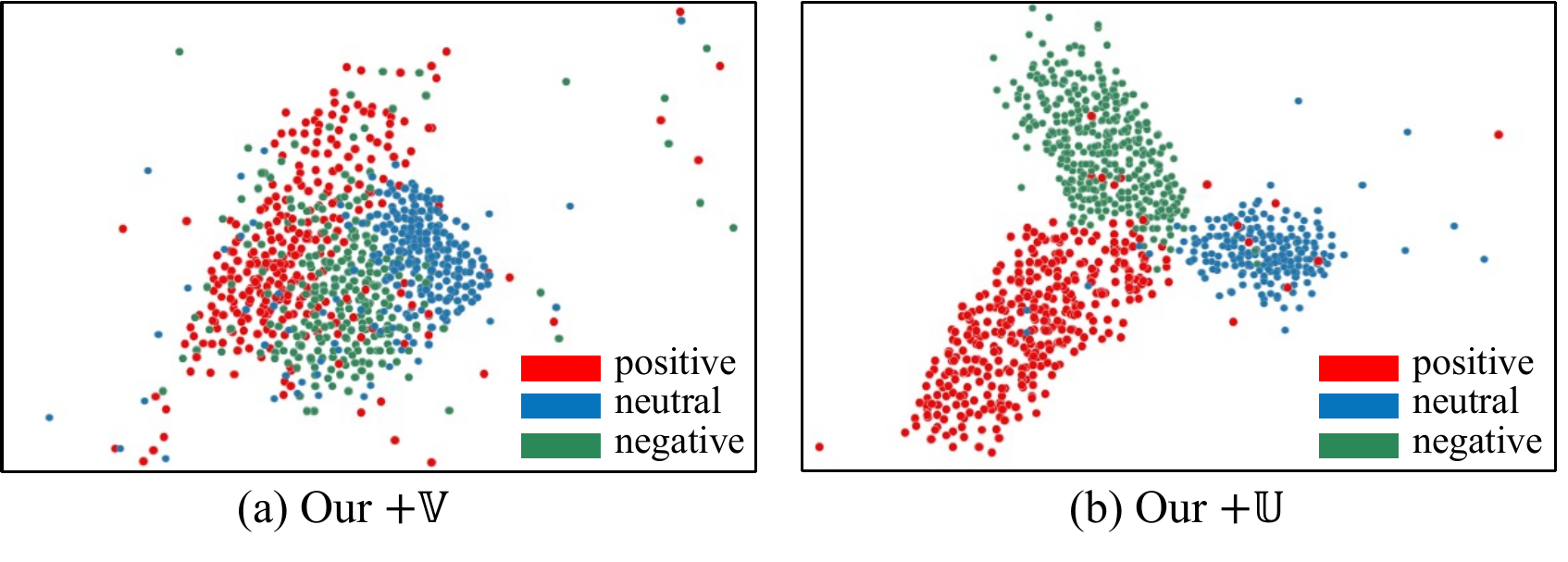}
\end{center}
\vspace{-0.4cm}
  \caption{Visualization of the features from two different latent space. We use the ViCo dataset~\cite{zhou2021responsive} for visualization. Three emotion are recorded in three different colors, there are 970 video clips of 2 seconds are collected.}
\label{fig compare ab}
\vspace{-0.2cm}
\end{figure}

\noindent \textbf{Analysis on the Latent Space Rearrangemet.} We conduct experiments to demonstrate the implications and importance of our latent space decomposition based on emotion priors. Specifically, as a comparison, we do not calculate the discrete latent space $\mathbb{U}$ through the emotional one-hot vector, but directly obtain the discrete latent space through the argument maximum on the \textit{Base Space}, which is called as $\mathbb{V}$ in the following part. The baseline methods NN-motion, NN-audio and DLS-Random are also repeated in $\mathbb{U}$ and $\mathbb{V}$ spaces, respectively. For a more intuitive understanding, we eval the facial coefficient ($\beta$) and head coefficient ($p$) in together. The ablation study results are shown in Table~\ref{table_all_ab} (we add the metrics value corresponding to $\beta$ and $p$ for comparison), we can find that $\mathbb{U}$ space has a stable improvement in each metric compared to $\mathbb{V}$ space, some outcomes are affected by randomness (RPCC on DLS-Rand.+$\mathbb{V}$ or $\mathbb{U}$). Then we visualize the results with $\mathbb{V}$ space and $\mathbb{U}$ space separately in Figure~\ref{fig compare ab face}. From the Figure~\ref{fig compare ab face} above, we can observe that in the $\mathbb{U}$ space, the positive listener exhibits a more pronounced grining, while the listener in $\mathbb{V}$ space only smiles slightly. On the other hand, from the below of Figure~\ref{fig compare ab face}, the negative listener in $\mathbb{U}$ space tends to show a more serious expression with more obvious frown. Furthermore, we adopt the t-SNE~\cite{van2008visualizing} to visualize the features decoded from the $\mathbb{U}$ and $\mathbb{V}$ spaces and take the corresponding emotion value as label. The emotional features in the $\mathbb{V}$ space (in Figure~\ref{fig compare ab}(a)) appear to be poorly decomposed, with the feature corresponding to different emotions being coupled together. While the $\mathbb{U}$ space (in Figure~\ref{fig compare ab}(b)) provides better feature distinction, as the value ranges of different emotions vary. \par

\begin{table}[t]
\footnotesize
% \vspace{-0.25cm}
\begin{center}
\setlength{\tabcolsep}{0.55mm}{
\begin{tabular}{lcccccccc}
\hlinew{1.15pt}
\multirow{3}{*}{} &\multicolumn{6}{c}{\textit{Coeff.} ($\beta+p$)} & & \multicolumn{1}{c}{Blink ($\phi$)}\\
\cline{2-7} \cline{9-9} %\cline{16-16}
&FD$\downarrow$ &V-D &SI-D &RPCC$\downarrow$ &WTLCC &STS$\downarrow$ & & WTLCC$\uparrow$\\
~&($\times 10^3$)& & & & ($\rightarrow$GT)  & ($\times 10^3$) & &   \\
\hline
\rowcolor{gtgray} GT &-- & 3.02 & 3.74 & -- & 0.329 & -- &  &--\\
\rowcolor{gray2}L2L~\cite{ng2022learning} &4.36 & 2.63  & 4.30 & \textbf{0.02} & 0.134 &13.40  & &  $1e^{-2}$\\
H=1 &7.93 &1.92  & 1.12 & 0.14 &0.122  &18.27 &  &0.37\\
H=4 &7.42 &3.15  & 2.38 & 0.14 & 0.171 & 11.34  &  &0.35\\
H=16 &4.14 & 2.27 & 2.55  & 0.11 & 0.193 & 10.81  & &0.38\\
H=64 & 3.67 & 3.70 & 3.52  & 0.12 & 0.228 & 9.72  & &0.41\\
H=128 &1.73 &3.29 &3.75 &0.09  &\textbf{0.312}&\textbf{7.02}&  &\textbf{0.42}\\
H=256 &\textbf{1.35} & 1.49 &1.26 &0.11  &0.370&7.35&  &0.42\\
\hlinew{1.15pt}
\end{tabular}}
\end{center}
\vspace{-0.1cm}
\caption{Ablation study results on the size of latent space are revealed by gradually increasing the value of H from 1 to 256. The L2L~\cite{ng2022learning} in this table is to compare with the results of H=1. We perform this experiment on the L2L~\cite{ng2022learning} datasets.}
\label{table_ab_2}
\end{table}

\noindent \textbf{Analysis on the size of Classification Head.} We discuss the impact from the size of the discrete space $\mathbb{U}$. With the definition of $\mathbb{U}$, it is evident that the size of configurations is $\rm{NV}^{H}$, which allows a vast motion space with a relatively small number of categories V. We set the default value of V to $64$ and varied the size of H to evaluate the effect on performance. It is worth noting that when we set H=$1$, our ASE is equivalent to the VQ-VAE~\cite{mirsamadi2017automatic} used in L2L~\cite{ng2022learning}, with the codebook size of V. The results are shown in Table~\ref{table_ab_2}, when H=1, our method exhibits slightly lower performance compared to L2L~\cite{ng2022learning}, as L2L~\cite{ng2022learning} utilizes a larger number of codewords than our default number of categories V. And the listener motion diversity and synchronicity increased (H from $1$ to $128$) with increasing latent space size. However, when H is excessively large ($256$), the performance of the model may be adversely affected, since the resulting latent space may be too large and hard to learn, ultimately leading to the difficulties in achieving an appropriate fit. Based on the aforementioned considerations, we set H to $128$.\par 

\begin{table}[t]
\small
\begin{center}
\setlength{\tabcolsep}{0.45mm}{
\begin{tabular}{cccccccccccc}
\hlinew{1.15pt}
\multirow{2}{*}{Method}& \multicolumn{2}{c}{Facial Expression} & & \multicolumn{2}{c}{Head Pose}&  &\multicolumn{2}{c}{Blink} &  &\multicolumn{2}{c}{All Aspect} \\
\cline{2-3} \cline{5-6} \cline{8-9} \cline{11-12}
~&Real  &Mesh & &Real &Mesh& &Real   &Mesh  & &Real   &Mesh   \\
\hline
RLHG~\cite{zhou2021responsive} & 1.4 & 1.2    & & 1.5  &1.2 & & 1.6 & 1.0  & & 1.5 & 1.1  \\
PCH~\cite{huang2022perceptual} &1.4 & 1.1     & & 1.4  &1.2 & & 1.1 & 1.1 & & 1.3 & 1.1  \\
L2L~\cite{ng2022learning} &3.7 & 2.3     & & 3.1 &2.3 & & 1.0 & 1.0 & & 3.3 & 2.9  \\
\rowcolor{gtgray} GT &4.6 & 4.9     & & 3.2  &3.9 & & 4.7 & 4.5  & & 4.9 & 4.6 \\
\rowcolor{mygray} Our &3.9 & 4.2     & & 4.7  &4.9 & & 4.3 & 4.5  & & 3.7 & 4.1 \\
\hlinew{1.15pt}
\end{tabular}}
\end{center}
\vspace{-0.1cm}
\caption{User study of our method with others. We calculate the average of the 5-scale scores for users. The ``Real'' means the photorealistic videos and ``Mesh'' means the mesh rendered videos. The facial and head movements can be evaluated more intuitively from the mesh videos.}
\label{table_user}
\end{table}

\subsection{User Studies}
\label{User Studies}

We conduct user studies to compare the generated results from the human perspective evaluation. We collected questionnaires from 42 users through an online platform. The questionnaire includes the generated videos and ground truth under different emotion communication situations. To eliminate any facial texture interference caused by the renderer, we also used the corresponding mesh video to evaluate the user perception of the synthesized listeners. Each user was asked to answer a Likert-type scale with the following options for all videos~\cite{kim2018deep}, ``Do you think the [$\times$] in video is like a listener?" (1-disagree, 2-weakly disagree, 3-normal, 4-weakly agree,5-agree), where the ``[$\times$]'' is for four criteria evaluated: 1) facial expression, 2) head pose movements, 3) blink and 4) all aspects. We record the average score for each type. Usually, in the experiment of 5-scale score, we think that more than $4$ points are close to the real performance. From the Table~\ref{table_user}, it can be found that our method has made great progress compared to the state-of-the-art methods, especially for the evaluation of mesh video and blink criteria. However, there is still a gap between our synthesized videos and ground truth videos, which is mainly due to the limitations of the resource-free renderer. \par

\section{Discussion and Reflections}
\label{Discussion and Reflections}

\begin{figure}[t]
\begin{center}
\includegraphics[width=0.9\linewidth] {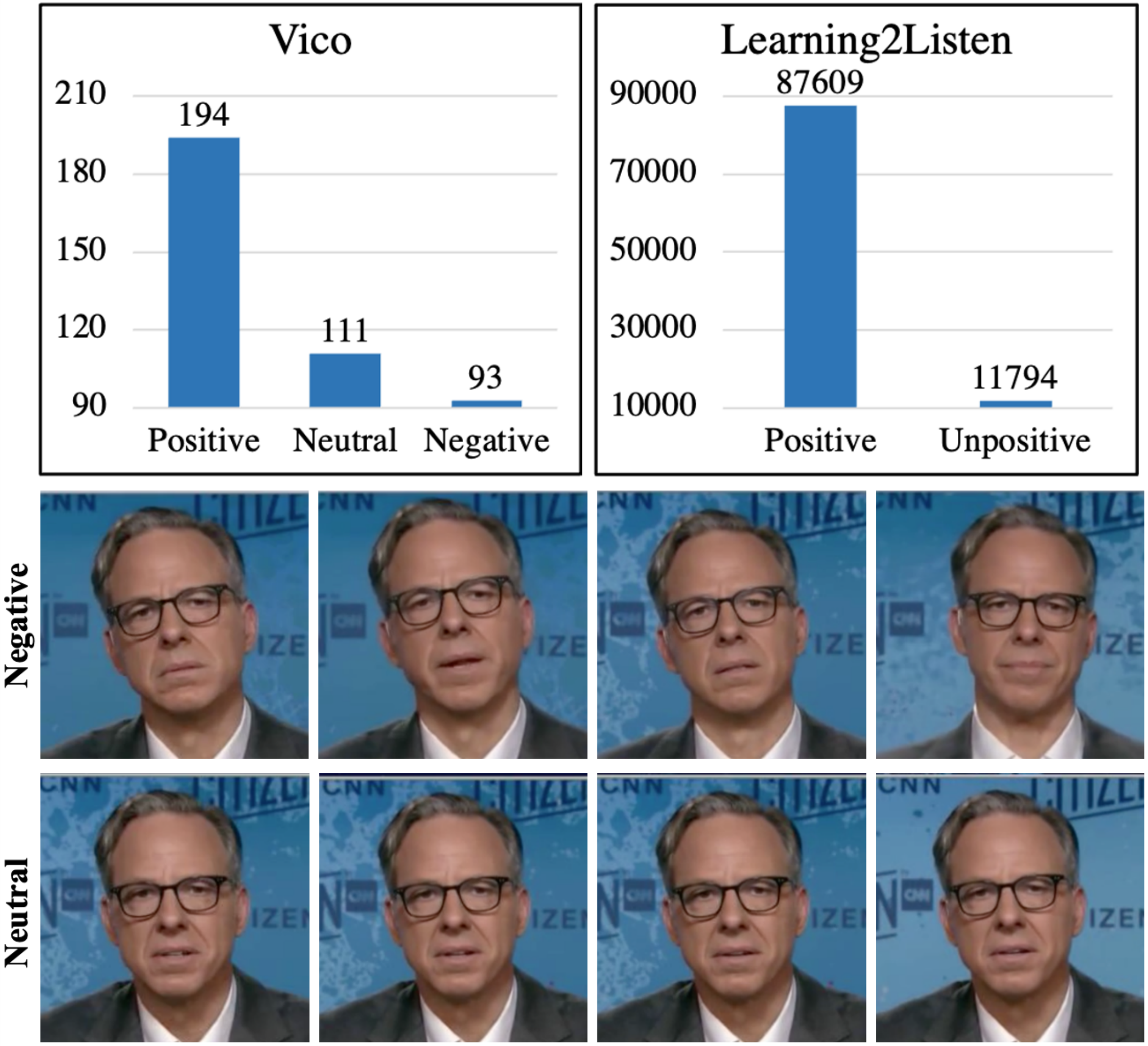}
\end{center}
\vspace{-0.1cm}
  \caption{The visualization of unbalance emotions and confusion annotation. The above shows the different emotion distributions in the RLHG~\cite{zhou2021responsive} and Learning2Listen~\cite{ng2022learning} datasets. For both datasets, the positive emotion is much higher than the other. We select the neutral and negative emotion from the ViCo~\cite{zhou2021responsive} dataset\protect\footnotemark, and visualize two different videos (each video represents one emotion). The emotions are not obvious.
}
\label{fig_datasets}
\vspace{-0.1cm}
\end{figure}
\footnotetext{The neutral video id is $@$JohCZ6VlC70\_000350\_000358 in ViCo, the negative video id is $@$JohCZ6VlC70\_000309\_000324 in ViCo.}

In this work, we present a novel listener motion synthesis method that takes into account the emotion in the dynamic conversation videos. To our best knowledge, our approach is the first exploration of emotion space representation for the listener. Although our method has proven to be remarkably efficacious under the current conditions, there are still some challenges. Two of the most tricky obstacles are (1) the emotional labels of listener in the dataset are unbalanced, and (2) the distance between emotions is unclear, that are visualized in Figure~\ref{fig_datasets}. In the Learning2Listen~\cite{ng2022learning} dataset, the videos with positive emotion are about eight times more than the the negative ones, the source of the videos makes this unbalanced (the entertainment interview videos from Youtube). The distance between negative emotions and neutral emotions in the ViCo dataset~\cite{zhou2021responsive} is not obvious, and it is even difficult to distinguish manually. We hope an extensive and sufficient listener dataset can be explored in the future. 

%-------------------------------------------------------------------------

%------------------------------------------------------------------------

{\small
\bibliographystyle{ieee_fullname}
\bibliography{egbib}
}

\end{document}